\documentclass[manuscript]{aastex}
\usepackage{amsmath}

\newcommand{\teff}{$T_{e\!f\!f}$}
\newcommand{\kms}{km s$^{-1}$}
\newcommand{\logg}{$\log g$} 

\shorttitle{The origin of the Aquarius Stream}
\shortauthors{Wylie de Boer et al.}

\begin{document}

\title{Tracing the origin of the Aquarius Stream: I}

\author{Elizabeth Wylie-de Boer and Kenneth Freeman}
\affil{Research School of Astronomy and Astrophysics, Australian National University, \\ Mount Stromlo Observatory, Cotter Rd, Weston Creek, ACT 2611, Australia}
    \email{ewylie@mso.anu.edu.au, kcf@mso.anu.edu.au}
\author{Mary Williams and Matthias Steinmetz}
\affil{Leibniz-Institut fuer Astrophysik Potsdam (AIP), \\An der Sternwarte 16, D-14482 Potsdam, Germany} \email{mary@aip.de}
\author{Ulisse Munari}
\affil{INAF Astronomical Observatory of Padova, 36012 Asiago (VI), Italy}
\and
\author{Stefan Keller}
\affil{Research School of Astronomy and Astrophysics, Australian National University, \\ Mount Stromlo Observatory, Cotter Rd, Weston Creek, ACT 2611, Australia}

\begin{abstract}
We present an abundance analysis of six member stars of the recently discovered Aquarius stream, in an attempt to ascertain whether this halo stream is real and, if so, to understand its origin. The mean metallicities of the six stars have a dispersion of only $0.10$ dex, indicating that they are part of a chemically coherent structure. We then investigate whether the stream represents the debris of a disrupted dwarf galaxy or a disrupted globular cluster.  The [Ni/Fe] - [Na/Fe] plane provides a good diagnostic: globular cluster stars and dwarf spheroidal galaxy stars are well separated in this plane, and the Aquarius stream stars lie unambiguously in the globular cluster region.  The Aquarius stream stars also lie on the  distinct [Na/Fe] - [O/Fe] and [Mg/Fe] - [Al/Fe] relations delineated by Galactic globular cluster stars. Spectroscopic parameters for the six Aquarius stars show that they are tightly confined to a $12$ Gyr, [Fe/H]$ = -1.0$, alpha-enhanced  isochrone, consistent with their identification as globular cluster debris. We present evidence that the Aquarius stream may continue through the disk and out into the northern halo. Our results indicate a globular cluster origin for the Aquarius stream, and demonstrate the potential for chemical tagging to identify the origins of Galactic substructures. \end{abstract}


\section{Introduction}
The idea of hierarchical galaxy formation, in which galaxies are believed to form from the aggregation of smaller elements was described by \citet{WhiteRees78}. In a related study from an observational viewpoint, \citet{Searle78} proposed that the halo of our Galaxy may have formed from the accretion of many small fragments.  Relics of these assembly processes may be recognizable as spatial, kinematic and chemical substructures among the Galactic stars.  The identification of debris from these ancient disrupted building-blocks is an important challenge for near-field cosmology, but is made more complex by the likely presence of debris from other old disrupted aggregates like globular clusters: see for example \citet{Gnedin}, \citet{Fall} and \citet{Odenkirchen}.  

Recently, Williams et al. (2011) identified 15 members of the newly discovered stellar Aquarius stream in data from the 2011 internal release of the RAdial Velocity Experiment (RAVE) (see \citet{RAVE} for the 3rd public data release). All 15 members have galactic longitudes and latitudes at $30^\circ < l < 75^\circ$ and $ -70^\circ < b < -50^\circ$, distances of up to about $10$ kpc and heliocentric line of sight velocities of around -200 km s$^{-1}$. The Aquarius stream appears to be a structure in the Galactic halo. From the RAVE stellar parameters, the metallicity distribution function and isochrone fits for the Aquarius stream stars are consistent with a 10 Gyr old population with metallicity [M/H] $\sim -1.0$.  In this paper, we present the first results of a more detailed element abundance study of Aquarius stream stars. Our goal is to determine whether the stream is real, and if so to ascertain whether the Aquarius stream stars are the debris of a disrupted dwarf galaxy, or a globular cluster.  Either outcome is interesting.  Globular cluster debris can give us insight into the cluster disruption processes and can serve as useful dynamical tracers for a variety of Galactic dynamics problems, even if the parent cluster has been completely disrupted.

\section{Data and Analysis}
The sample of six Aquarius stream stars was taken from Williams et al. (2011). Spectra were acquired with the echelle spectrograph on the 2.3m telescope at Siding Spring Observatory over the period 2010 July to 2011 August. The spectrograph provides complete wavelength coverage from 4120 to 6920\AA\ with a resolution of about 25,000. Our spectra have an average signal to noise per resolution element of about 30.  Bias frames, quartz lamp exposures for flat-fielding and Th-Ar exposures for wavelength-calibration were all included.  Bright radial velocity standards and hot rapidly rotating stars for telluric line removal were also observed each night with much higher signal to noise. The data were reduced using standard IRAF routines in the packages \textbf{imred}, \textbf{echelle} and \textbf{ccdred} with no deviation from the normal reduction procedures.

Radial velocities for the whole Williams sample of Aquarius stars were independently
measured with the Asiago echelle spectrograph, and confirm  that the mean velocity of the Williams sample is correct to about 1 \kms, and the random errors are about 2.0 \kms.  

\subsection{Abundance Analysis}\label{abunds}
Spectroscopic atmospheric parameters were derived for each star using a chi-squared fitting program which compares observed spectra to a model grid of synthetic spectra from \citet{Munari05}.  The program takes several regions of the spectra and matches them against $\sim 7000$ synthetic spectra covering the range \teff\ $ = 4000$ to $6000$ K, \logg\ $ = 0.0 $ to $5.0$, [M/H] $ = -2.5$ to $+0.5$ with [$\alpha$/Fe] $ = 0.0$.  The results are shown in Table \ref{params}.  The final atmospheric parameters were obtained by averaging over the values from the different regions, and the $\pm$ value quoted in Table \ref{params} is the standard deviation over the values from the different regions.  The regions used were 5575 - 5725\AA, identified as being sensitive to effective temperature, and 4950 - 5200\AA, identified as being sensitive to both surface gravity and metallicity.  An initial [$\alpha$/Fe] $ = 0.0$ was assumed as it was not intended to unnecessarily bias the results in any way prior to the abundance analysis.  Before the abundance analysis was undertaken there was no reason to believe the [$\alpha$/Fe] value in for these stars would be any different from scaled solar.  Microturbulence values, $\xi$, were obtained using established relationships from \citet{Reddy03} for dwarf stars (Equation 1) and \citet{Kirby08} for giant stars (Equation 2).  These values and uncertainties are also included in Table \ref{params}.  The uncertainty shown in $\xi$ is calculated using the associated uncertainties on \teff\ and \logg\ as well as the corresponding term's constant from the equation and then adding these in quadrature.  For the subgiant, C2309161-120812, the microturbulence was found by averaging the results given by both Equation 1 and 2.  

\begin{equation}
\xi = (1.28 + 3.3 \times 10^{-4}\times(T_{eff} - 6000) - 0.64 \times (log g - 4.5))\ km\ s^{-1}
\end{equation}

\begin{equation}
\xi = (2.70 - 0.509\times log g)\ km\ s^{-1}
\end{equation}

The average absolute difference between RAVE parameters and those obtained in this study were $\Delta$\teff\ $ = 159$ K, $\Delta$\logg\ $ = 0.48$ and $\Delta$[M/H]$ = 0.34$ dex. Our mean errors are $136$ K in \teff, $0.22$ in \logg, $0.18$ dex in [M/H] and 0.18 km s$^{-1}$ in $\xi$. The effect of these uncertainties on derived abundances is discussed further in Section \ref{uncabund}.  

Abundances for individual elements were obtained via the method of spectrum synthesis using the MOOG code \citep{Sneden73} and \citet{Anders89} solar abundances.  The analysis for these stars included atomic line lists from
 Kurucz\footnote{http://kurucz.harvard.edu} with refined oscillator
strengths for all lines not involving hyperfine splitting via a reverse solar analysis using the Kitt Peak solar spectrum\footnote{http://bass2000.obspm.fr/solar$_{-}$spect.php}.  This reverse solar analysis was done with a solar model of \teff\ $ = 5770$ K and \logg\ $ = 4.44$.  The linelist used for abundances can be found in Table \ref{linelist}.  For two hotter stars in which weaker Na lines could not be measured accurately enough (C2309161-120812 and J223811.4-104126), the NaD lines were used to obtain the sodium abundance.  Likewise, for the single star in which weaker Mg lines could not be measured (J223811.4-104126), the Mgb lines were used to obtain the magnesium abundance.  In such hotter metal-poors stars, these lines do not saturate and provide a reliable estimate of these abundance although with greater errors due to broadening.  The spectrum synthesis program MOOG has an option to deal with strong lines, such as NaD and Mgb, separately during the analysis.  Hyperfine splitting components were taken into account for the lines of Sc, V and Co.  

\subsubsection{Abundance Uncertainties}\label{uncabund}
Generally, there are three main sources of uncertainty when measuring abundances: uncertainty due to atmospheric parameters, uncertainty in measurement of individual lines and the standard deviation in abundances from many lines of the same species.

Perhaps the major source of uncertainty in abundance comes from the uncertainty in the adopted atmospheric parameters for the stellar models. Table \ref{abundsens} shows the sensitivity of the log $\epsilon$ abundance values for representative errors in atmospheric parameters for these stars: $\pm$ 150K in effective temperature, $\pm$ 0.25 dex in gravity and $\pm$ 0.2km s$^{-1}$ in microturbulence (taken from Table \ref{params}).  These errors were measured by repeating the abundance determination for all lines used (listed in Table \ref{linelist}) with models of varying parameters.  When added in quadrature, the uncertainty due to associated errors in atmospheric parameters can be as much as +0.28 dex, although over the species and individual lines used in this analysis averages at +0.16 dex.  Every effort was made to minimise the effect of microturbulence by using weaker lines that are not as sensitive to errors in $\xi$.  In some cases the combined uncertainty of X and Fe in the value [X/Fe] cancels each other out, e.g. [Mg/Fe].  However, in cases such as [O/Fe] these errors do not cancel.  

In addition to the uncertainty due to atmospheric parameters, the one-dimensional, local thermal equilibrium (LTE) assumption currently made in atmospheric models also introduces an additional error source (see \citet{Asplund05} for review of 1D, non-LTE effects).  While the absolute value of these errors is currently unknown for most elements, large efforts are being put into both understanding and calculating these errors, as well as the creation of 3D, non-LTE atmospheric model grids for stars which will enhance the inclusion of non-LTE errors into abundance calculations \citep{Collet11}.

The spread in derived abundances from multiple lines of the same species is a further, but usually smaller, source of error.  This can be
due to incorrect $\log gf$ values, although in this study a reverse
analysis using the sun was done for all elements to reduce this
problem (see earlier in section for more details).  In Table \ref{abund}, $\sigma$ represents the standard deviation of abundances derived from multiple lines, and in this study is always less than 0.10 dex except in the case of [Ni/Fe].  

Where abundances are obtained from only one line (as for O, Na, Si and Ba and occasionally Al and V) a more subjective error estimate needs to be made, and the quoted $\sigma$-value is the estimated uncertainty from a single line measurement.  Although every line was analysed by hand and the associated abundance carefully determined to reduce this uncertainty, the resulting uncertainty can vary depending on line strength, blending and crowding in the region, and the overall signal to noise and quality of the data, which in this study was not high. Consequently, this error contribution is estimated to be about $0.05$ to $0.15$ dex.  

\section{Results and Discussion}

This work confirms the existence of Aquarius as a coherent structure based on its chemistry.   The six stars in our sample cover a metallicity range [M/H] $= -1.25$ to $-0.98$, with a mean 
[M/H] of $-1.15$ and standard deviation $\sigma = 0.10$. We can compare these values with those for the larger Williams et al. sample of $15$ stars (mean [M/H] $\sim -1.0$ and $\sigma = 0.44$): the means are consistent, but the abundance spread of our smaller sample is much smaller and consistent with no intrinsic spread. We note that the RAVE data have significantly lower resolution ($R = 7500$) than the data for this study. The [Fe/H] values for our study range from [Fe/H] $ = -1.15$ to $-0.93$, with a mean of $-1.09$ and $\sigma = 0.10$.  The final [Fe/H] value was obtained from 16 Fe I and 3 Fe II lines, as listed in Table \ref{linelist}.  The ionisation balance, or the agreement between abundances derived from Fe I and Fe II lines, is a good test of the adopted gravity.  Table \ref{abund} shows abundance obtained for both Fe I and Fe II.  The difference between these two species was never more than 0.05 dex, showing that the adopted gravities are acceptable.  The average absolute difference between our [Fe/H] values and the [M/H] values obtained from the initial chi-squared program (see Section \ref{abunds}) is $0.05$ dex (see Table 3).

\begin{figure}[htbp]
\plotone{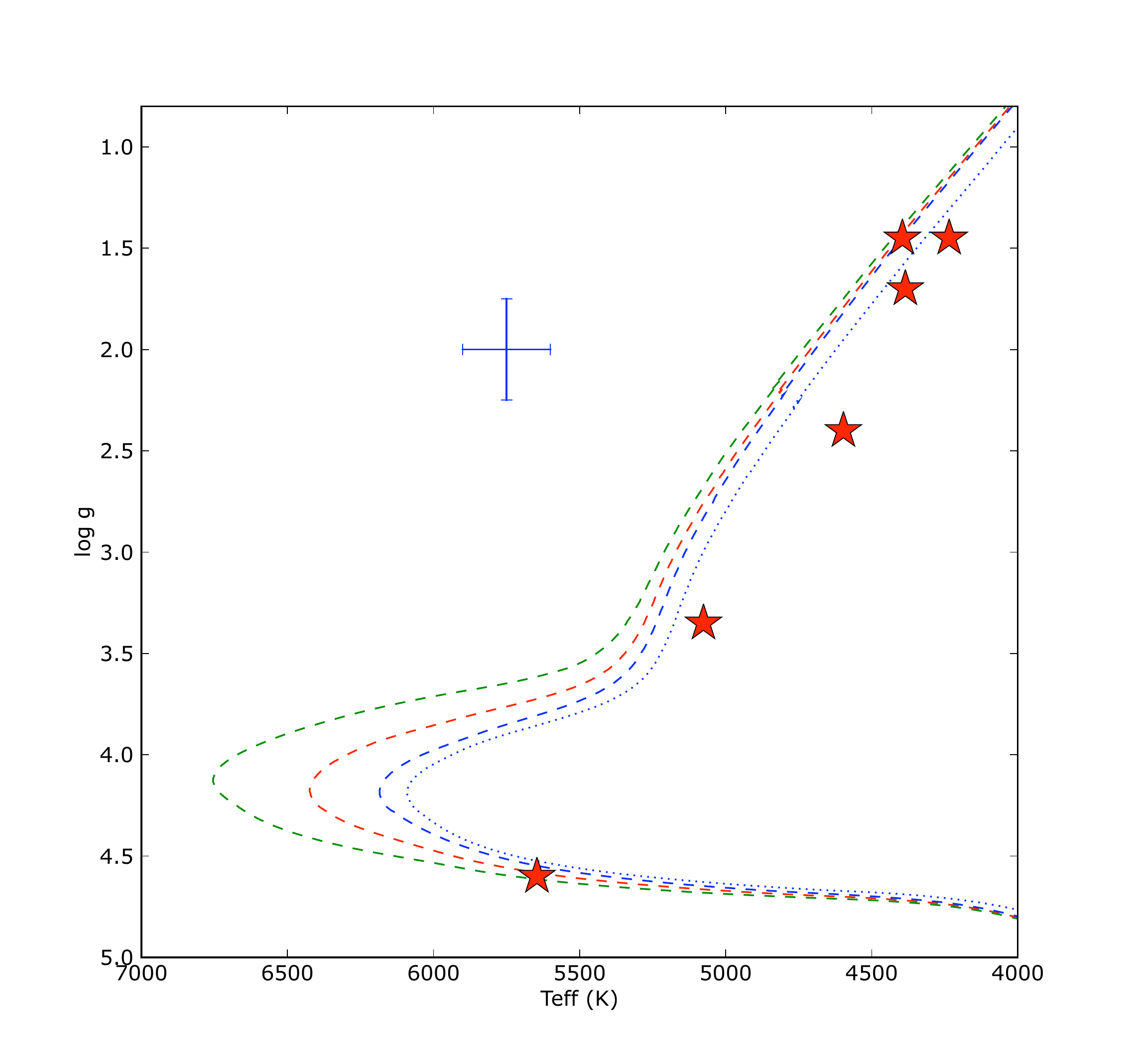}
\caption{Four Dartmouth Stellar Evolutionary Database isochrones \citep{Dotter07} for ages $=$ 12, 9 and 6 
Gyr, [Fe/H] $= -1.0$ and alpha-enhancements of [$\alpha$/Fe] $= +0.2$ (blue, red and green dashed lines respectively) and for age $= 12$ Gyr, [Fe/H] $= -1.0$ and  [$\alpha$/Fe] $= +0.4$ (blue dotted line), with the six Aquarius stars overplotted.  A typical error bar for this study is shown. }
\label{cmd}
\end{figure}

All six Aquarius stars show significant alpha-element enhancement (from Mg, Si, Ca and Ti).  The average enhancements are [Mg/Fe] $ = +0.30$, [Si/Fe] $ = +0.25$, [Ca/Fe] $ = +0.25$ and [Ti/Fe] $ = +0.24$.  Alpha enhancement seen at this metallicity implies that enrichment of this stellar population includes contribution from core-collapse supernovae. Relative to the high- and low-$\alpha$ populations observed in the Galactic halo by \citet{Nissen10}, our Aquarius stars belong to the old, high-$\alpha$ population, with one possible exception.  The six stars lie close to several alpha-enhanced isochrones in the \teff, \logg\ plane. Figure 
\ref{cmd} shows a 12, 9 and 6 Gyr isochrone for [Fe/H] $ = -1.0$ and [$\alpha$/Fe] $ = +0.2$ (blue, red and green dashed lines respectively) as well as a 12 Gyr, [Fe/H] $ = -1.0$ and [$\alpha$/Fe] $=+0.4$ (blue dotted line) from the Dartmouth Stellar Evolutionary Database isochrones \citep{Dotter07}.  As the current set of data contains no stars in the age-sensitive turn-off region, it is unclear if the younger (9 and 6 Gyr) isochrones fit more reasonably than the 12 Gyr isochrone adopted by \citet{Williams11}.  Therefore, in this current study, a younger age population can not be ruled out.  While the 12 Gyr isochrone appears to fit slightly better at the base of the giant branch, all we can do is confirm the conclusion made in \citet{Williams11} - that a 12 Gyr isochrone, corresponding to an old, metal-poor population, fits the data.  

\begin{figure}[htbp]
\plotone{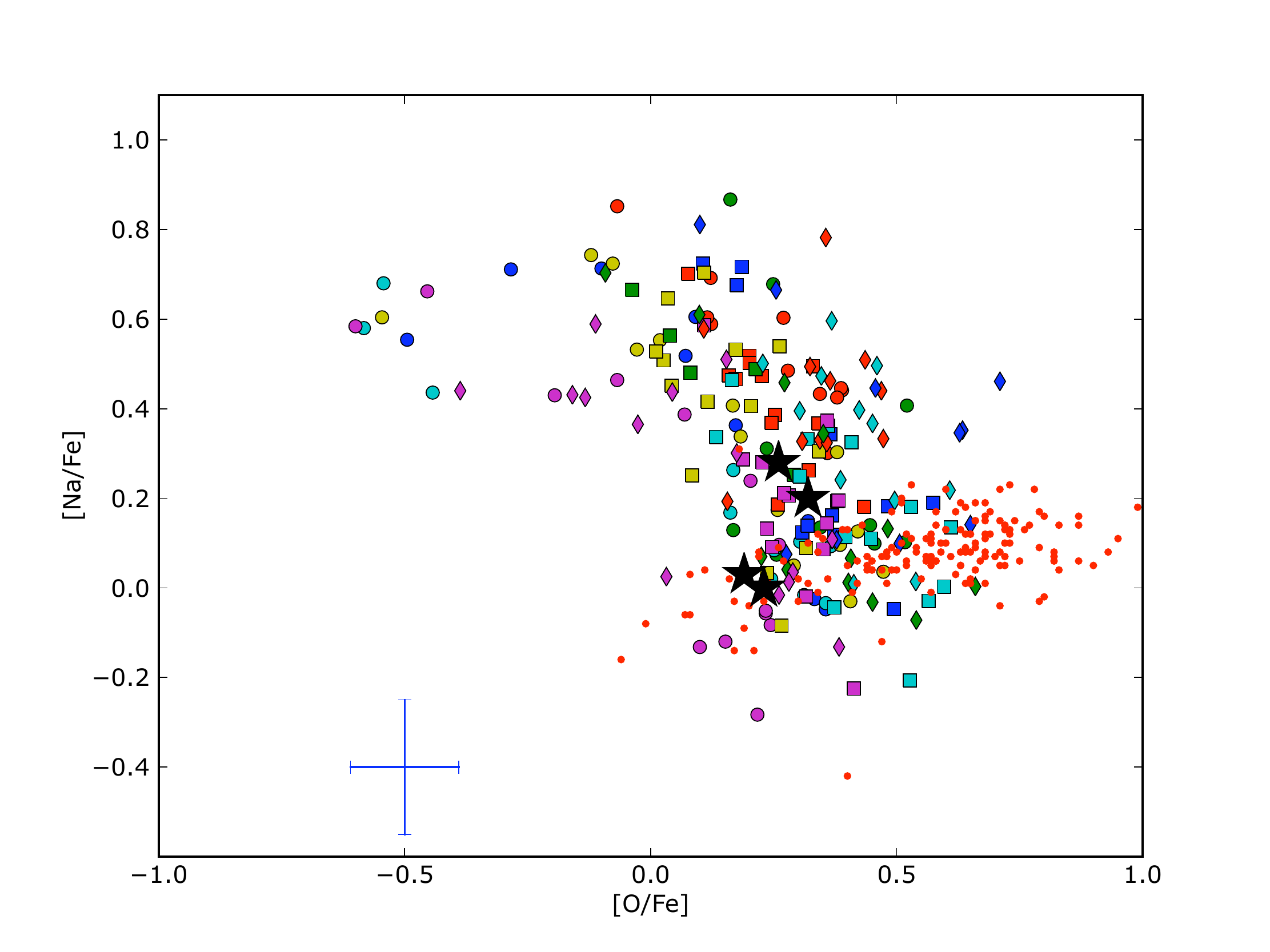}
\caption{The Na-O anticorrelation for globular clusters, taken from \citet{Carretta09nao}, where different colours and symbols represent the 19 different globular clusters studied, and Galactic field stars from \citet{Reddy06} and \citet{Fulbright00} (small red circles).  The four black stars represent the Aquarius stars for which both Na and O could be reliably measured.  A typical error bar for this study is shown.  }
\label{nao}
\end{figure}

\begin{figure}[htbp]
\plotone{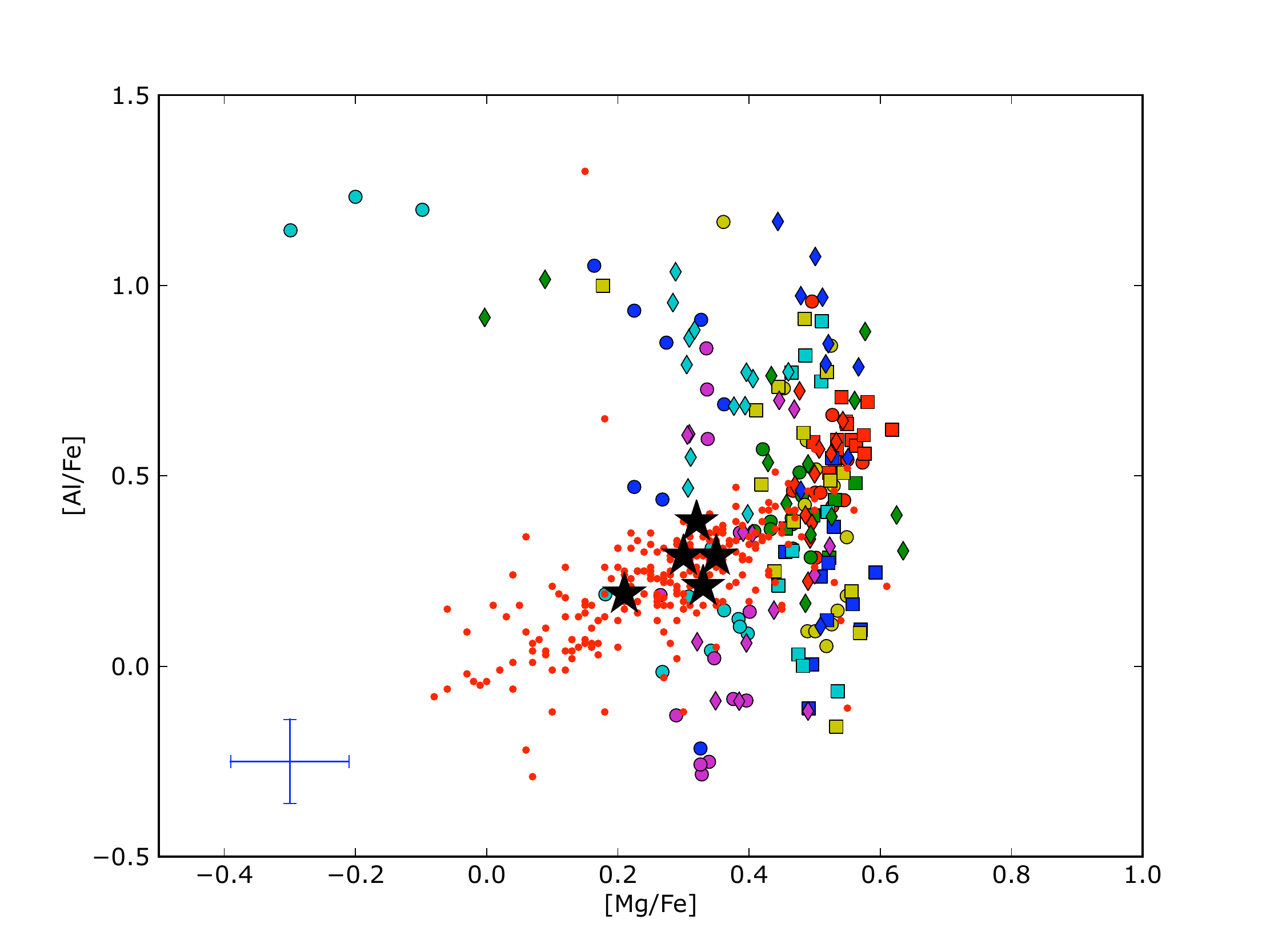}
\caption{The Mg-Al relationship for globular clusters and Galactic field stars.  The symbols are the same as in Figure \ref{nao}.  The five black stars represent the Aquarius stars for which both Mg and Al could be reliably measured.   A typical error bar for this study is shown. }
\label{mgal}
\end{figure}

In determining whether the Aquarius stream originated from a  globular cluster or a dwarf spheroidal galaxy, we can use the well-established relationships of Na-O and Mg-Al abundances in Galactic globular clusters as a guide. These relationships, from  
\citet{Carretta09nao,Carretta09mgal}, are shown in Figures \ref{nao} and \ref{mgal} respectively,  with different colours and symbols representing different globular clusters.  Galactic field stars from \citet{Reddy06} and \citet{Fulbright00} are also shown as small red circles. The Aquarius stream stars are shown as large black stars.  Only four Aquarius stream stars could be measured for Na-O and five for Mg-Al. For these stars at least, the Na-O and Mg-Al abundances lie close to those for the globular clusters and are consistent with a globular cluster origin. They do not
exclude a dwarf spheroidal galaxy origin, because Na-O and Mg-Al relationships have not yet been established for dwarf spheroidal galaxies: the O and Al lines are weak in these systems.

\begin{figure}[htbp]
\plotone{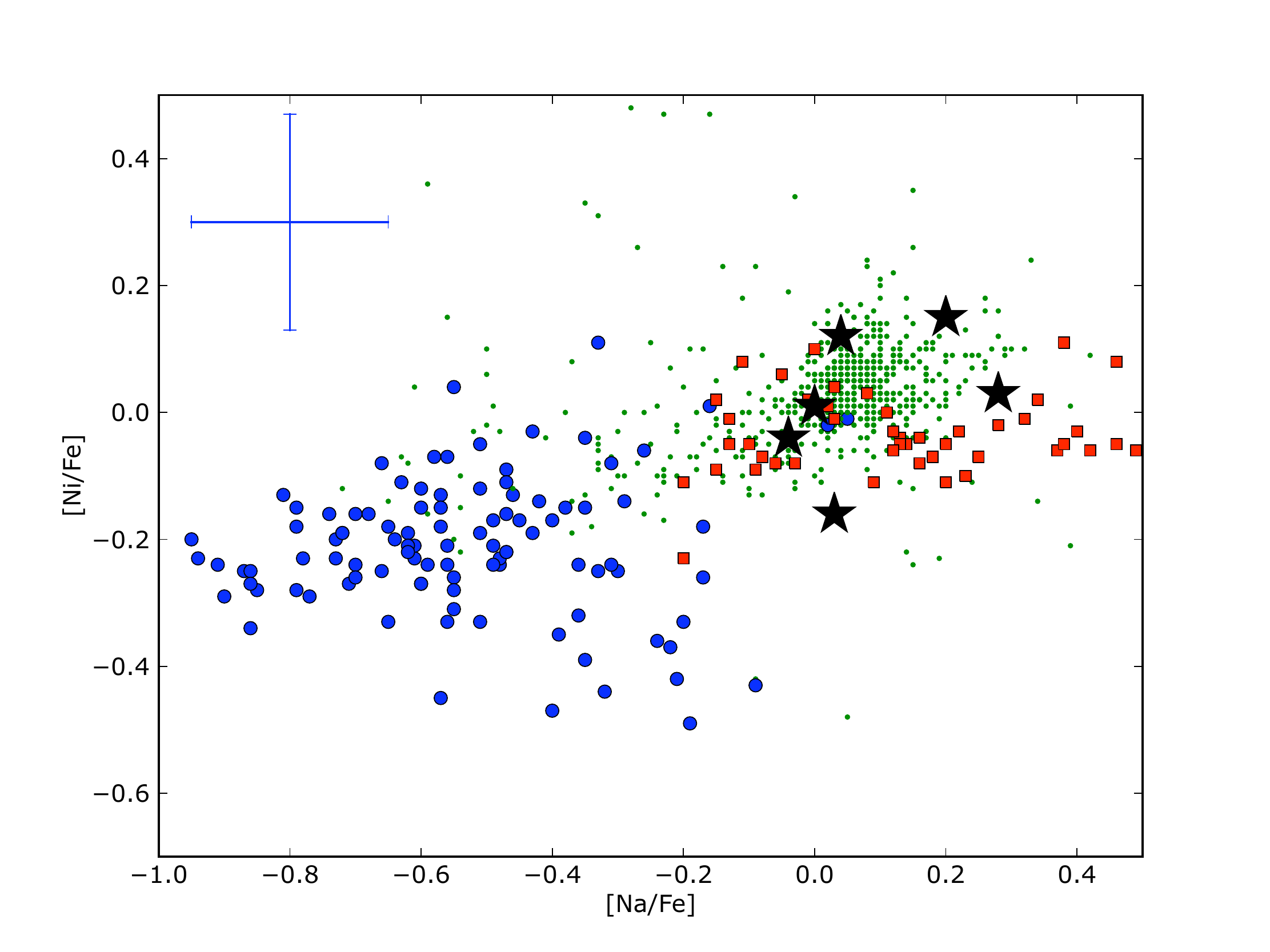}
\caption{The relationship between Na and Ni in dwarf spheroidal galaxies as blue circles (Fornax \citep{Letarte10}, Sagittarius \citep{Sbordone07}, Sculptor, Leo I and Carina \citep{Shetrone03}); Galactic stars as green points, \citep{Venn04}; globular clusters stars as red squares (NGC 288 and 362 \citep{Shetrone00} and M3 \citep{Sneden04} and the six Aquarius stream stars from this study as large black stars.  A typical error bar for this study is shown. } 
\label{nani}
\end{figure}

The relationship between  Na and Ni abundances as seen in Figure \ref{nani} provides a useful discriminant.  This relationship is discussed in more detail in \citet{Venn04}, \citet{Letarte10} and references therein.  In summary, a Na-Ni correlation is expected when the chemical enrichment is dominated by SNeII because the 
production of $^{23}$Na and $^{58}$Ni depend on the neutron excess.  Figure \ref{nani} shows the Na-Ni values for five nearby dwarf galaxies from various studies \citep{Letarte10, Sbordone07,Shetrone03}, globular cluster stars \citep{Sneden04,Shetrone00}, Galactic field stars \citep{Venn04} and our Aquarius stream stars.  The dwarf spheroidal galaxies display sub-solar Na and Ni values, in marked contrast to the globular clusters and Aquarius stream stars that show solar or mild Na-Ni enhancements more like those seen in the Galactic field stars.  One possible explanation would be that [Ni/Fe] and [Na/Fe] in dwarf spheroidal galaxy stars are shifted to lower values due to additional Type Ia SNe production of Fe, which is not relevant in globular cluster environments.  However, this does not agree with nucleosynthesis calculations (eg. \citet{Nomoto97}), which predict that Ni is overproduced relative to Fe in Type Ia SNe.  While this discrepancy is not easily explained, it seems to be well established observationally.  A shift in the [Ni/Fe] - [Na/Fe] relation of low-alpha halo stars relative to high-alpha stars is also seen by \citet{Nissen10}.  Figure \ref{nani} indicates that the Aquarius stream stars did not originate in a dwarf spheroidal galaxy, but are more likely the debris of a disrupted globular cluster.
  
\section {Morphology of the Aquarius Stream}

To interpret the nature of the Aquarius stream, \citet{Williams11} made simulations 
of a evolution of a dissolving small galaxy orbiting in the Galactic potential. They estimated 
the tangential motion of the stream by using the (reduced proper motion) - (J-K) diagram 
for the stream stars. The distribution of stars in this plane is a good match to their 
isochrone for an adopted tangential velocity $v_{T}$ of 250 km s$^{-1}$ . We note that the Aquarius stream covers a fairly small area of sky and has stars over a wide range of distances: it appears to be coming almost directly towards us. If this is correct, then its tangential velocity $v_T$ would come mostly from the projected $v_T$ of the sun in the direction of the stream, which is about $215$ \kms. A stream orbit with this value of $v_T$ would still be consistent with the reduced proper motion distribution, within the errors.

For their simulation, Williams et al. modelled the dissolution of a small galaxy with 
initial core radius of 300 pc and velocity dispersion 10 km s$^{-1}$ . They found a good match 
to the distribution and kinematics of the stream stars after 700 Myr from the start of the 
simulation, and inferred that the stream was dynamically relatively young and not phase 
mixed. If the Aquarius stream is in fact the debris of a globular cluster, for which the 
core radius and velocity dispersion could have been as small as 2 pc and 2 km s$^{-1}$ , then a 
somewhat longer dynamical age may be consistent with the distribution and kinematics.

The six stars observed here, and the other 9 stars identified by Williams et al. as Aquarius Stream members, lie in a small range of galactic longitude and latitude on the south side of the Galactic disk ($30^\circ < l < 75^\circ$ and $ -70^\circ < b < -50^\circ$).  Their distances extend from a few hundred pc to several kpc, and the stream therefore appears to be passing close to the solar neighborhood. The mean velocities of the stream stars are more negative for the closer stars, decreasing from about $-160$ \kms\ at a distance of $3$ kpc to about $-210$ \kms\ at $1$ kpc.  The stream appears to be plunging in towards the solar neighborhood from the south, and is likely to extend out to the north on the opposite side of the Galactic disk, i.e. into the region $210^\circ < l < 255^\circ$, $ 50^\circ < b < 70^\circ$.  If the stream does extend to the north, we can make a rough estimate of the likely velocities of the northern stream stars by assuming that the stream kinematics are antisymmetric north-south within a few kpc of the Galactic plane.  From the data of Williams et al (2011), the velocities of northern stream stars would then lie in a range from about $160$ to $240$ \kms.  

The southern Aquarius stars identified from the RAVE survey are all relatively bright ($K < 11.3$), and the stream was detected at high contrast against the sparse background of bright halo stars.  We would therefore like to search for similarly bright northern stream stars. The RAVE survey does not extend into this potential northern stream region.  We searched various other sources of relatively bright  metal-poor stars, and found potential northern stream stars among the brighter stars in the SDSS survey and also in the Chiba \& Beers (2000) paper (and references therein).  

\begin{figure}[htbp]
\plotone{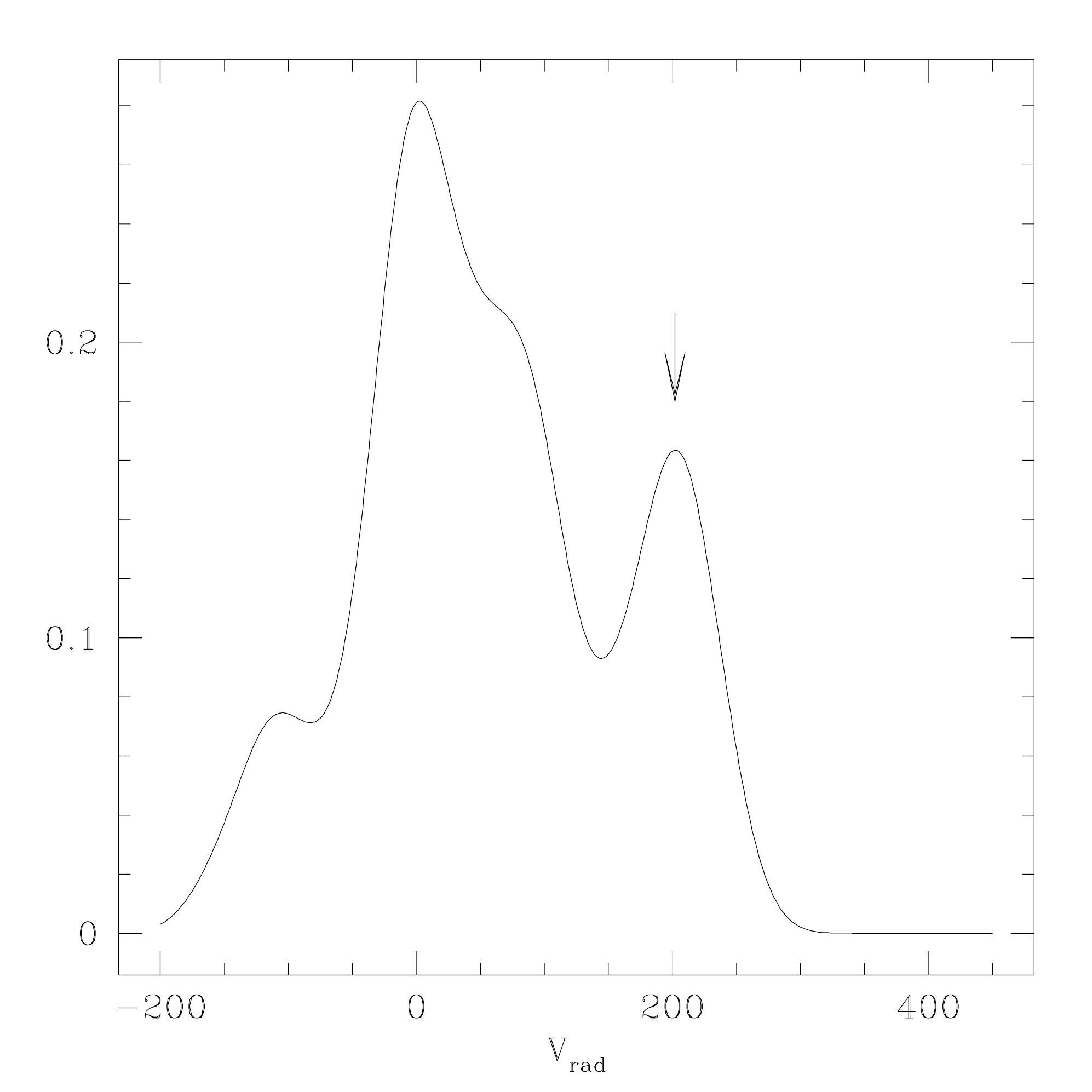}
\caption{The velocity distribution of 61 northern SDSS stars selected in the range of [Fe/H] and galactic longitude and latitude consistent with membership of a northern extension of the Aquarius stream. The distribution is shown as a generalized histogram with smoothing kernel width $\sigma = 25$ \kms. The arrow shows a strong secondary peak in the distribution at $202$ km s$^{-1}$. }
\label{sdss}
\end{figure}

For the SDSS search, the SDSS DR7 was used  \citep{SDSS}.  We looked for stars in the above range of $(l,b)$ with $-1.0 > {\rm [Fe/H]} > -1.7$, magnitudes in the range $14.7 < g < 16.0$, colors $0.2 < g-r < 1.0$ and velocity errors $< 20$ km s$^{-1}$. The SDSS stars are fainter than the RAVE stars, and the faint magnitude limit at $g = 16.0$ was chosen to reduce the background of non-stream halo stars which increases rapidly for fainter magnitudes.  Figure \ref{sdss} shows the velocity distribution of these 61 SDSS stars, presented as a generalized histogram with a gaussian smoothing kernel ($\sigma = 25$ km s$^{-1}$) to avoid binning and to reduce the effect of small sample noise. (Each star is represented by a gaussian distribution centered at its observed velocity.) The arrow shows a strong secondary peak in the velocity distribution at $202$ km s$^{-1}$), close to the mean of the expected velocity for the northern extension of the Aquarius stream as estimated above. For comparison, the radial velocity associated with the projected solar motion in the center of the northern stream field is $88$ \kms. 

The \citet{Chiba} compilation includes one RR Lyrae star, ST Leo, whose position on the sky, radial velocity and metallicity are consistent with membership of the northern Aquarius stream.  It also includes three RR Lyrae stars which are potential members of the (southern) Aquarius stream on the basis of their position, abundance and velocity.  The positions, [Fe/H] values, heliocentric radial velocities and mean V magnitudes for the four RR Lyrae stars are given in Table 4.  Their [Fe/H] values go from $-1.1$ to $-1.6$; the two RR Lyrae stars with high resolution abundance measurements have abundances of $-1.18$ and $-1.08$. The six stars described in this paper have [Fe/H] in the range $-0.91$ to $-1.18$.  We conclude that (a) our six stars and the two RR Lyrae stars with high resolution abundance data are consistent with being part of a coherent structure with a homogeneous globular-cluster-like MDF (mean [Fe/H] $= -1.10$ and observed dispersion $= 0.09$ dex), and (b) the Aquarius stream appears to extend across the Galactic disk into the northern halo, and that it is likely to be the debris of a disrupted globular cluster.  

\section{Conclusion}
Our abundance study of six Aquarius stream stars supports the reality of the stream discovered by \citet{Williams11}. From the stellar parameters given in Table 1, the Aquarius stream is a chemically coherent structure that appears to be populated along its length from a distance of at least 5 kpc (the two brightest giants) to 150 pc from the sun (the dwarf). The existing sample of Aquarius stream stars, which were identified from RAVE stars in a fairly narrow range of K magnitude between about $9.5$ and $11$, must therefore represent only a small fraction of the stream stars. Large numbers of Aquarius stream stars remain to be discovered, including stars in its likely northern extension.

The stream appears to be a relatively narrow feature, passing through the disk near the solar neighborhood and out into the halo on the other side. With a sample of accurate distances and radial velocities for stream tracers, the stream could provide stringent constraints on the total gravitational field along the stream and hence on the ratio of stellar to dark matter in the disk near the sun. Accurate proper motions would enhance these constraints.  The RR Lyrae stars would be ideal tracers.  Table 4 lists four bright RR Lyrae stars that are possible stream members.  A survey for fainter RR Lyrae stars in the stream fields could provide accurate tracers of the stream to the south and north out to much larger distances from the sun than the six stars discussed here.

The element abundances of the Aquarius stream stars demonstrate again the power of chemical tagging to identify the origin of stars which were born together in a now-disrupted cluster or dwarf galaxy. This study, and previous studies such as those by \citet{Desilva}, \citet{Bubar} and \citet{me}, have applied chemical tagging techniques to determine the nature of substructures which were already identified kinematically, such as the Aquarius stream and the HR1614, Wolf 630 and Kapteyn moving groups. \citet{Karlsson} were able to identify stars from a disrupted cluster in the Sextans dwarf spheroidal galaxy using chemical tagging techniques, without prior kinematical indicators of membership.  The upcoming GALactic Archaeology with HERMES survey (GALAH), using the new HERMES\footnote{www.aao.gov.au/HERMES} instrument on the AAT, will go a step further in Galactic reconstruction. The primary goal of this high resolution spectroscopic survey of a million Galactic stars is to use chemical tagging techniques to identify substructures which have lost their kinematical and spatial identity through disruption, phase mixing and heating events.
 
\acknowledgments

We are grateful for the use of the ANU 2.3m telescope, to the technical staff at Siding Spring Observatory who keep it going, to Massimo Fiorucci for the use of his $\chi^2$ program and to Amanda Karakas for discussions.  We would like to thank the referee for their suggestions to improve the paper.  This work is supported by an ARC Discovery program grant DP0772283 and DP120104416.

\clearpage

\begin{deluxetable}{ccccccccccc}
\tablewidth{0pt}
\rotate
\tablecaption {Stellar parameters for the six Aquarius stars, as derived by RAVE ($_R$) and this analysis\label{params}}
\tablehead{
\colhead{ID}& \colhead{J\tablenotemark{a}} & \colhead{K\tablenotemark{a}} & \colhead{V\tablenotemark{b}$_{los}$} &\colhead{\teff$_{R}$\tablenotemark{b}}&  \colhead{log g$_{R}$\tablenotemark{b}} &\colhead{$[$M/H$]_{R}$\tablenotemark{b} }&\colhead{\teff} & \colhead{log g} &\colhead{ $[$M/H$]$} &\colhead{$\xi$}\\
 \colhead{}& \colhead{}& \colhead{}& \colhead{km s$^{-1}$}&\colhead{K }&\colhead{ }&\colhead{ }&\colhead{K} &\colhead{} &\colhead{}&\colhead{km s$^{-1}$}}
\startdata
C2225316-145437 & 10.34 & 9.57 & -155.7&4104 &1.01 & -1.29 & 4235 $\pm$118 & 1.45 $\pm$0.21 & -1.20 $\pm$0.14 & 1.96 $\pm$ 0.11 \\
C2309161-120812 & 10.68 & 9.97 & -224.1&5219 & 2.94 & -0.66 & 5076 $\pm$127 & 3.35 $\pm$0.35  & -1.10 $\pm$0.14 & 1.35 $\pm$ 0.23\\
J221821.2-183424 & 10.34 & 9.68 & -154.1&4572 & 1.06 & -1.54 & 4395 $\pm$205 & 1.45 $\pm$0.35  & -1.15 $\pm$0.21 & 1.96 $\pm$ 0.18\\
J223504.3-152834 & 10.36 & 9.65 & -166.9&4795 & 3.05 & -0.33 & 4597 $\pm$158 & 2.40 $\pm$0.14 & -0.98 $\pm$0.17 & 1.47 $\pm$ 0.07\\
J223811.4-104126 & 10.42 & 9.90 & -230.1&5502 & 4.16 & -0.78 & 5646 $\pm$147 & 4.60 $\pm$0.15 & -1.20 $\pm$0.20 & 1.09 $\pm$ 0.11  \\
J232619.4-080808 & 10.51 & 9.76 & -218.7 & 4225 & 1.14 & -1.22 & 4385 $\pm$ 61 & 1.70 $\pm$ 0.14 & -1.25 $\pm$ 0.21 & 1.80 $\pm$  0.07\\
\enddata
\tablenotetext{a}{From 2MASS catalogue}
\tablenotetext{b}{From \citet{Williams11}}
\end{deluxetable}

\clearpage 

\begin{deluxetable}{cccc}
\tablewidth{0pt}
\tablecaption {Linelist used for abundance analysis\label{linelist}}
\tablehead{
\colhead{Species} &\colhead{ Wavelength} & \colhead{Ex Pot} & \colhead{log gf}
}
\startdata
O I & 6300.30 & 0.00 & -9.82 \\
Na I  & 5889.95 & 0.00 & 0.12 \\ 
Na I & 5895.92 & 0.00 & -0.18 \\
Na I & 6154.23 & 2.10 & -1.56 \\
Na I & 6160.75 & 2.10 & -1.26 \\
Mg I & 5172.68 & 2.71 & -0.40 \\
Mg I & 5183.60 & 2.72 & -0.18 \\
Mg I & 5711.09 & 4.34 & -1.83 \\
Mg I & 6318.72 & 5.10& -1.73 \\
Mg I & 6319.24 & 5.10 & -1.95 \\
Al I & 6696.02 & 3.14 & -1.35 \\
Al I & 6698.67 & 3.14 & -1.65\\
Si I	&	5948.54 & 5.078 & -1.23 \\
Ca I	&	6102.72 & 1.878 & -0.89 \\ 
Ca I	&	6122.22 & 1.884 & -0.41 \\
Ca I	&	6162.17 & 1.897 & 0.10 \\
Ca I	&	6439.08 & 2.524 & 0.47 \\
Ca I	&	6449.81 & 2.519 & -0.55 \\
Ca I 	&	6455.60 & 2.521 & -1.35 \\
Ca I	&	6471.66 & 2.524 & -0.59 \\
Ca I	&	6493.78 & 2.519 & 0.14 \\ 
Sc II	&	5641.001 \tablenotemark{a}& 1.499 & \\
Sc II 	&	5657.90 \tablenotemark{a}& 1.506 & \\
Sc II	&	6279.56 \tablenotemark{a} &1.499 &  \\
Sc II	&	6604.60 \tablenotemark{a}& 1.356 & \\
Ti I	&	6126.22 & 1.066 & -1.43 \\
Ti I	&	6258.10 & 1.442 & -0.35 \\
Ti I    &      6258.71 & 1.459 & -0.24 \\
Ti I 	&	6261.10 & 1.429 & -0.48 \\
V I	&	6251.80 \tablenotemark{a} & 0.286 &  \\
V I	&	6268.80 \tablenotemark{a} & 0.286 & \\
V I	&	6274.20  \tablenotemark{a} &2.113 & \\
Cr I	&	5788.39 & 3.011 & -1.49 \\
Cr I	&	6330.09 & 0.941 & -2.92 \\
Fe I &   6056.00      &     4.730  &   -0.46              \\                         
Fe I &   6065.48      &     2.607    & -1.53               \\                        
Fe I &   6078.49      &     4.792     &-0.48                \\                        
Fe I &   6079.01      &     4.648 &    -1.12                  \\                      
Fe I &  6136.62      &     2.452    & -1.40        \\                                
Fe I &   6136.99      &     2.196    & -2.95         \\
Fe I &     6173.33      &     2.221 &    -2.88         \\                               
Fe I &   6344.15      &     2.431   &  -2.92             \\                           
Fe I &   6358.70      &     0.858   &  -4.47               \\                         
Fe I &   6392.54      &     2.277   &  -4.03                 \\                      
Fe I &   6393.60      &     2.431   &  -1.62                  \\                      
Fe I &   6408.02      &     3.684   &  -0.97   \\                                     
Fe I &   6411.65      &     3.651   &  -0.82     \\                                   
Fe I &   6420.06      &     4.577   &  -2.73       \\                                 
Fe I &   6421.35      &     2.277   &  -2.03         \\                               
Fe I &   6430.85      &     2.174   &  -2.01           \\                             
Fe II &   6149.26      &     3.886   &  -2.72       \\
Fe II &   6416.92      &     3.889   &  -2.74         \\                               
Fe II &   6432.68      &     2.889 &    -3.71           \\                             
Co I	&	6282.60  \tablenotemark{a} &1.739 & \\
Co I   &	6450.10  \tablenotemark{a} & 2.135& \\
Co I	&	6454.80  \tablenotemark{a} &3.629 & \\
Ni I	&	6314.65 & 1.934 & -1.77 \\
Ni I	&	6482.80 & 1.934 & -2.63 \\ 
Ni I	&	6643.63 & 1.675 & -2.30 \\
Ba II	&	6496.90 & 0.604 & -0.407 \\
\enddata
\tablenotetext{a}{Hyperfine splitting included}
\end{deluxetable}

\clearpage
\begin{deluxetable}{ccccc}
\tablewidth{0pt}
\rotate
\tablecaption{Sensitivity of log $\epsilon$ abundance value to atmospheric parameters\label{abundsens}}
\tablehead{
\colhead{Species }& \colhead{$\Delta$ Teff }&\colhead{$\Delta$ log g } &\colhead{$\Delta$ $\xi$ } &\colhead{Total} \\
\colhead{}&\colhead{+150K}&\colhead{+0.25 dex}&\colhead{+0.2 km s$^{-1}$}&\colhead{\tiny{added in quadrature}}
}
\startdata
Fe	&	0.13	&	0.03	         &	-0.10 	&	0.17	\\
O	&	0.03	&	0.11  	&	0.00 	&	0.11	\\
Na	&	0.14	&	0.00 	&	-0.06 	&	0.15	\\
Mg	&	0.09	&	0.00 	&	-0.03 	&	0.09	\\
Al	&	0.11	&	-0.02 	&	-0.01 	&	0.11	\\
Si	&	-0.03	&	0.08  	&	-0.05 	&	0.10	\\
Ca	&	0.17	&	-0.01 	&	-0.08 	&	0.19	\\
Sc	&	-0.07	&	0.13  	&	-0.04 	&	0.15	\\
Ti	&	0.25	&	-0.02 	&	-0.02 	&	0.25	\\
V	&	0.28	&	-0.02 	&	-0.01 	&	0.28	\\
Cr	&	0.21	&	-0.01 	&	-0.01 	&	0.21	\\
Co	&	0.14	&	0.04	         &	-0.06 	&	0.15	\\
Ni	&	0.14	&	0.06  	&	-0.07 	&	0.17	\\
Ba	&	0.05	&	0.02  	&	-0.08 	&	0.10	\\
\enddata
\end{deluxetable}

\begin{deluxetable}{ccccccccccccc}
\tablewidth{0pt}
\rotate
\tablecaption{Average abundances and uncertainties\label{abund}}
\tablehead{
\colhead{Species }& \colhead{C2225316}&\colhead{ $\sigma$ }&\colhead{ C2309161} &\colhead{ $\sigma$ }& \colhead{J221821} &\colhead{$\sigma$} &  \colhead{J223504} &\colhead{$\sigma$ }& \colhead{J223811} &\colhead{ $\sigma$} & \colhead{J232619} & \colhead{$\sigma$}
}
\startdata
$[$M/H$]$     & -1.20	& 0.14 & -1.10 & 0.14  & -1.15 & 0.21 & -0.98 & 0.17 & -1.20 & 0.20 & -1.25 & 0.21\\ 
$[$Fe I/H$]$    & -1.18	& 0.12 & -1.04 & 0.09  & -1.14 & 0.06 & -0.91 & 0.09 & -1.14 & 0.11 & -1.17 & 0.09\\ 
$[$Fe II/H$]$    & -1.13	& 0.11 & -1.08 & 0.12  & -1.10 & 0.07 & -0.96 & 0.10 & -1.10 & 0.10 & -1.13 & 0.10\\ 
$[$O/Fe$]$    & 0.23 	& 0.12 & ...         & ...     & 0.19  & 0.12  & 0.26  & 0.12 & ...        & ...      & 0.32 & 0.12 \\
$[$Na/Fe$]$  & 0.00 & 0.15 & -0.04\tablenotemark{a} & 0.15 & 0.03 & 0.15 & 0.28 & 0.21 & 0.04\tablenotemark{a} & 0.15 & 0.20 & 0.04\\
$[$Mg/Fe$]$  & 0.32 & 0.05  & 0.21 	& 0.05 & 0.30 & 0.01 & 0.33  & 0.05 & 0.30\tablenotemark{b} & 0.06 & 0.35 & 0.10\\
$[$Al/Fe$]$    & 0.38 & 0.12  & 0.19	&  0.07       & 0.29 & 0.12       & 0.21  & 0.03 & ...       &...        & 0.29 & 0.09\\
$[$Si/Fe$]$    & 0.26 & 0.10  & 0.09	&0.10      & 0.24 & 0.10 & 0.33  & 0.10 & 0.28  & 0.10   & 0.32 & 0.10\\
$[$Ca/Fe$]$  & 0.25 & 0.08  & 0.21	& 0.05 & 0.21 & 0.08& 0.31  & 0.04 & 0.24  & 0.10   & 0.27 & 0.05\\
$[$Sc/Fe$]$  & 0.02  & 0.04  & -0.03 	& 0.03 & 0.00 & 0.02& 0.08  & 0.11 & ...  & ...   & 0.09 & 0.06\\
$[$Ti/Fe$]$   & 0.28  & 0.02  & 0.22 	& 0.03 & 0.16 & 0.06& 0.23  & 0.03 & ...  & ...   & 0.29 & 0.11\\
$[$V/Fe$]$    & 0.00  & 0.03  & ... 	& ... & -0.09& 0.15     & 0.11 & 0.11 & ...      & ...       & 0.09 & 0.03\\
$[$Cr/Fe$]$  & -0.02  & 0.10  & 0.02 	& 0.04 & ...      & ... & 0.06  & 0.10 & 0.14      & 0.12       & 0.09 & 0.05\\
$[$Co/Fe$]$ & -0.01  & 0.02   & -0.06 & 0.12 & -0.14 & 0.23 & 0.07  & 0.08 & ...      & ...       & -0.01 & 0.05\\
$[$Ni/Fe$]$  & 0.01  & 0.06   & -0.04 	& 0.02 & -0.16 & 0.21& 0.03  & 0.03 & 0.12 & 0.18 & 0.15 & 0.23\\
$[$Ba/Fe$]$  & 0.03 & 0.10   & 0.06 	& 0.10 & 0.13 & 0.10  & 0.03  & 0.10 & 0.24 & 0.10 & 0.22 & 0.10\\
\enddata
\tablenotetext{a}{Abundance derived solely from Na D lines}
\tablenotetext{b}{Abundance includes that derived from MgB lines}
\end{deluxetable}
\clearpage

\begin{deluxetable}{cccrcccc}
\tablewidth{0pt}
\rotate
\tablecaption {Four RR Lyrae stars: possible Aquarius stream members\label{RRL}}
\tablehead{
\colhead{ID} & \colhead{RA(2000)} & \colhead{Dec(2000)} & \colhead{{\it l}} & \colhead{{\it b}} & \colhead{[Fe/H]} & \colhead{$V_{los}$\tablenotemark{a}} & \colhead{V\tablenotemark{a}}\\
 \colhead{} & \colhead{} & \colhead{} & \colhead{} & \colhead{} & \colhead{} & \colhead{km s$^{-1}$} &\colhead{mag} 
}
\startdata
BV Aqr & $22~02~54.0$ & $-21~31~32$ & $32$  & $-51$ & $-1.18\tablenotemark{b}$ & $-239$ & $10.9$ \\
BN Aqr & $22~27~48.8$ & $-07~29~02$ & $56$  & $-51$ & $-1.33\tablenotemark{d}$ & $-182$ & $12.5$ \\
DN Aqr & $23~19~17.2$ & $-24~12~59$ & $36$  & $-69$ & $-1.63\tablenotemark{d}$ & $-220$ & $11.2$ \\
ST Leo & $11~38~32.7$ & $+10~33~42$ & $253$ & $+66$ & $-1.08\tablenotemark{c}$ & $+168$ & $11.5$ \\
\\
\enddata
\tablenotetext{a}{From \citet{Chiba}}
\tablenotetext{b}{From \citet{Solano} - high resolution}
\tablenotetext{c}{From \citet{Fernley} - high resolution}
\tablenotetext{d}{From \citet{Layden} - low resolution}

\end{deluxetable}


\end{document}